\documentclass{article}
\usepackage{spconf,amsmath,graphicx}
\usepackage{amssymb}
\usepackage{booktabs}
\usepackage{multirow}
\usepackage{hyperref}
\usepackage{enumitem}
\usepackage{xcolor}
\usepackage{pifont}

\newcommand{\cmark}{\ding{51}}%
\newcommand{\xmark}{\ding{55}}%

\title{DeCoAR 2.0: Deep Contextualized Acoustic Representations with Vector Quantization}
\name{Shaoshi Ling, Yuzong Liu}
\address{Amazon AWS AI\\
\normalsize \texttt{\{shaosl,liuyuzon\}@amazon.com}}
\begin{document}
\ninept
\maketitle
\begin{abstract}

Recent success in speech representation learning enables a new way to leverage unlabeled data to train speech recognition model. In speech representation learning, a large amount of unlabeled data is used in a self-supervised manner to learn a feature representation. Then a smaller amount of labeled data is used to train a downstream ASR system using the new feature representations. Based on our previous work DeCoAR \cite{ling2020deep} and inspirations from other speech representation learning, we propose DeCoAR 2.0, a \textbf{De}ep \textbf{Co}ntextualized \textbf{A}coustic \textbf{R}epresentation with vector quantization. We introduce several modifications over the DeCoAR: first, we use Transformers in encoding module instead of LSTMs; second, we introduce a vector quantization layer between encoder and reconstruction modules; third, we propose an objective that combines the reconstructive loss with vector quantization diversity loss to train speech representations. Our experiments show consistent improvements over other speech representations in different data-sparse scenarios. Without fine-tuning, a light-weight ASR model trained on 10 hours of LibriSpeech labeled data with DeCoAR 2.0 features outperforms the model trained on the full 960-hour dataset with filterbank features. 


\end{abstract}

\begin{keywords}
speech recognition, acoustic representation learning, semi-supervised learning
\end{keywords}

\section{Introduction}
\label{sec:intro}


In the long history of semi-supervised learning (SSL) in speech recognition, self-training approach \cite{vesely2013semi,huang2016semi,synnaeve2019end} and knowledge distillation \cite{hinton2015distilling}, or known as teacher-student model training \cite{li2014learning} are the two commonly used SSL methods. Recent success of representation learning enables a new approach towards leveraging unlabeled data. In natural language processing community,  BERT~\cite{devlin2019bert}, ELMo~\cite{peters2018deep}, XLNet \cite{yang2019xlnet}, GPT  \cite{radford2018improving} and its follow-ups are classical examples of representation learning. The key philosophy of representation learning is based on using self-supervised learning, where we obtain `free' labels from unlabeled data and train them in a supervised manner via some proxy tasks. In the context of BERT~\cite{devlin2019bert}, two proxy tasks are defined including masked language model task and two-sequence prediction task. These proxy tasks are designed to force the learning of a robust, meaningful representation.  After the representation has been learned, a downstream task model is then trained using labeled data with the learned representation. Optionally, the representation learning block and downstream task block can be fine-tuned together. 

Learning efficient speech representation can be traced back to restricted Boltzmann machine \cite{hinton2006fast, hinton2012deep, bengio2007greedy}, which allows pre-training on large amounts of unlabeled data before training the deep neural network speech models.  More recently, speech representation learning has drawn increasing attention in speech processing community and has shown promising results in semi-supervised speech recognition \cite{schneider2019wav2vec, baevski2020wav2vec, ling2020deep, liu2020tera}.  The design of proxy tasks in learning speech representation can be categorized into two types. The first type is based on \textbf{contrastive loss}~\cite{van2017neural} and has been applied to speech representation such as wav2vec and its variants \cite{schneider2019wav2vec,baevski2019vq,baevski2020wav2vec}. The model is trained to learn representations containing information that most discriminates the future or masked frame from a set of negative samples via contrastive loss.  The second type is based on \textbf{reconstructive loss}. The proxy task for these representation learning methods is to reconstruct temporal slices of acoustic features based on contextual information. These reconstruction tasks can be defined as autoregressive reconstruction, or masked-based reconstruction. APC \cite{chung2019unsupervised} and its follow-up \cite{chung2020improved} are examples to use autoregressive reconstruction loss.  In many state-of-the-art pretrained language model task, masked-based prediction is adopted in the proxy tasks such as BERT \cite{devlin2019bert} and XLNet \cite{yang2019xlnet}.  In speech, instead of prediction, we randomly mask temporal slices of acoustic features and attempt to reconstruct them \cite{jiang2019improving, song2019speech, wang2020unsupervised, liu2020mockingjay, chi2020audio, liu2020tera}.

Orthogonal to the contrastive-/reconstructive-loss based speech representation learning, vector-quantized speech representations have been proposed~\cite{van2017neural, liu2020towards, baevski2019vq, baevski2020wav2vec, chung2020vector}. One motivation to apply vector quantization (VQ) is that enforcing quantization can lead to better linguistic unit discovery \cite{oord2018representation, harwath2019learning} due to the discrete nature of phonetic units. In VQ-APC \cite{chung2020vector}, the authors use VQ as a way to limit model capacity and control information needed in encoding representation. In VQ-wav2vec \cite{baevski2019vq} and wav2vec 2.0 \cite{baevski2020wav2vec}, the author use VQ to facilitate direct application of BERT and other NLP algorithms.

In this paper, we introduce DeCoAR 2.0, a \textbf{De}ep \textbf{Co}ntextualized \textbf{A}coustic \textbf{R}epresentation with vector quantization. We take inspirations from many recent advances in speech representation learning, and propose multiple improvements over vanilla DeCoAR. We summarize the contributions of this paper as follows:
\begin{itemize}[leftmargin=*,itemsep=0pt, topsep=1pt]
    \item We propose to use Transformer as encoding block and replace LSTM in the vanilla DeCoAR;
    \item We present a deep contextualized acoustic representation learning approach with the addition of a vector quantization layer;
    \item We propose a new objective function that combines masked-based reconstruction loss with VQ diversity loss.
\end{itemize}

\section{Related Work}
\label{sec:related}

\subsection{An Overview on DeCoAR}

\begin{figure}[th]
  \centering
  \includegraphics[width=0.9\linewidth]{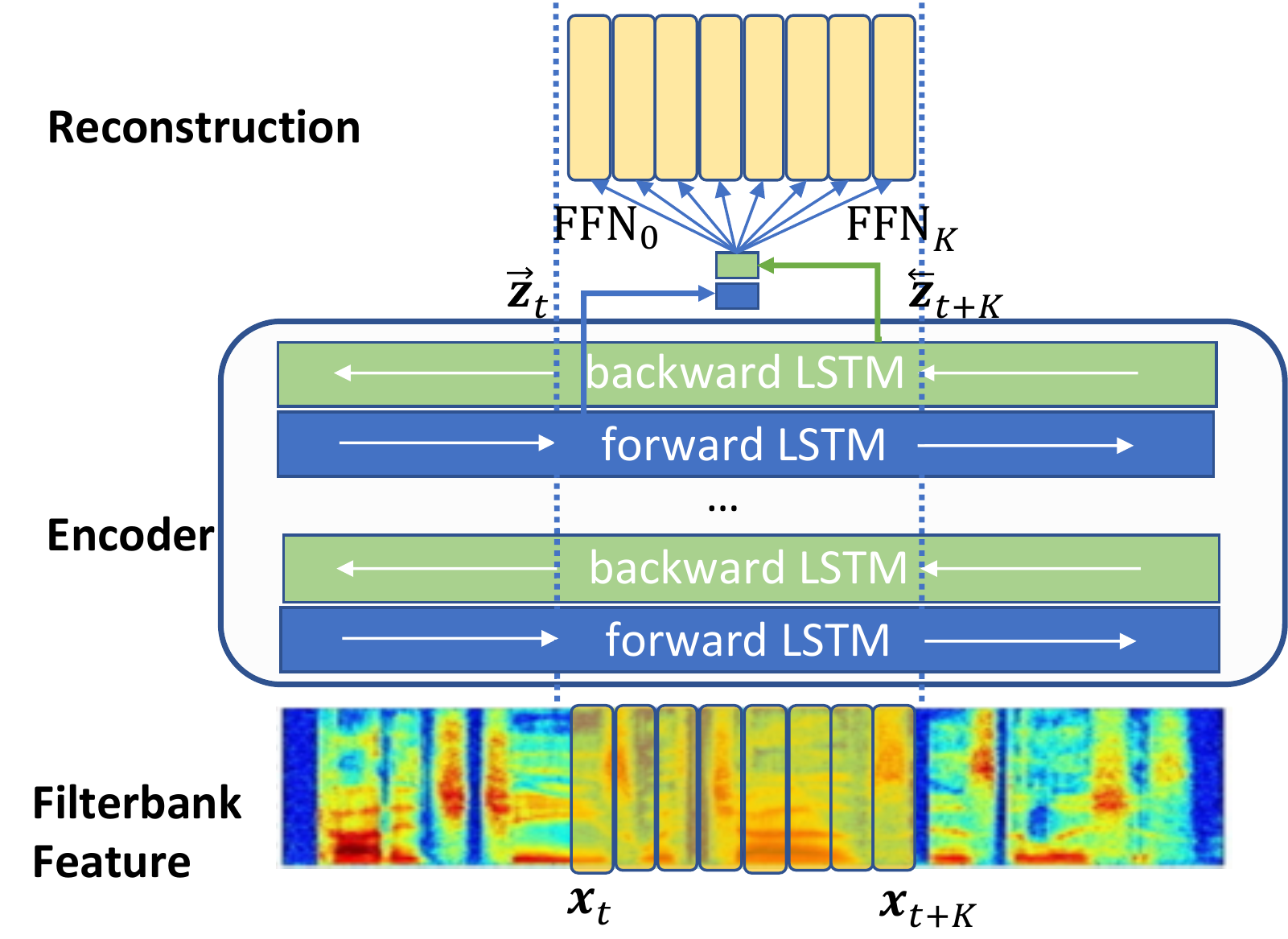}
  \caption{Illustration of DeCoAR.}
  \label{fig:decoar}
\end{figure}

DeCoAR stands for \textit{deep contextualized acoustic representations}, and was proposed in our previous work~\cite{ling2020deep}. As depicted in Figure~\ref{fig:decoar}, DeCoAR consists of two modules, an encoder module and a reconstruction module. For an input speech sequence $\mathbf{X} = (\mathbf{x}_1,\cdots,\mathbf{x}_T)$, an \textbf{encoder module} consists of a stacked forward and backward LSTMs, and computes a hidden representation that encodes information from both previous and future frames (i.e. $\overrightarrow{\mathbf{z}}_t, \overleftarrow{\mathbf{z}}_t$). For each temporal slice $(\mathbf{x}_t, \mathbf{x}_{t+1}, ..., \mathbf{x}_{t+K})$, the \textbf{reconstruction module} takes the concatenated forward state at time $t$ and backward state at $t+K$ as inputs, and uses  position-dependent feed-forward networks to recontruct each frame. Formally, the DeCoAR objective is defined as follows:
\begin{equation}\label{eq:decoar}
  \begin{aligned}
\mathcal{L}_t= \sum_{i=0}^{K}|\mathbf{x}_{t+i} - \text{FFN}_i ([\overrightarrow{\mathbf{z}}_t; \overleftarrow{\mathbf{z}}_{t+K}])|
\end{aligned}
\end{equation}
where $\text{FFN}_i$ is a position-dependent feed-forward network to reconstruct the $i$-th frame in the slice.  The final loss $\mathcal{L}$ is calculated over all possible slices in the entire sequence in an autoregressive manner, defined as: $\mathcal{L}=\sum_{t=1}^{T-K}\mathcal{L}_t$.

\subsection{Vector-quantized Representation Learning}
\subsubsection{wav2vec 2.0}
Wav2vec 2.0 \cite{baevski2020wav2vec} is one of the successful examples in representation learning. It uses 10 minutes of labeled data with 53k hours of unlabeled data to achieve a word error rate (WER) of 5.2\%/8.6\% on LibriSpeech benchmark. The model relies on a diverse codebook learned to correlate the underlying speech units to representations via contrastive loss. Discretizing the continuous representation enables applications of many state-of-the-art NLP algorithms. In wav2vec 2.0, after applying VQ operations, the model is trained using a masked LM style loss, similar to BERT. 

One potential challenge in learning optimal codebooks with contrastive loss is posed by data with nuisance factors such as noise and other adverse conditions. In these cases, the codebook can be trivially optimized by assigning acoustic condition (e.g. voice activity, noise) to the codebook. A potential work-around is to use frame reconstruction as objective so that the network can leverage all available information of the input feature to guide the learning of a robust representation. 


\subsubsection{VQ-APC}
VQ-APC \cite{chung2020vector} introduced an novel approach that inserted a VQ layer before frame prediction. The motivation of using VQ is to quantify the information needed to encode speech representation and control the capacity of the models. The model uses autoregressive predictive coding (APC) as objective, instead of a contrastive predictive coding (CPC). Their experiments showed APC/reconstruction objective performed better than CPC/constrastive objective under the same condition. They also demonstrated the learned VQ codes highly correlate to phoneme path, suggesting VQ can be used to capture linguistic units in an implicit way. 

\begin{figure*}[thb]
  \centering
  \includegraphics[width=0.8\linewidth]{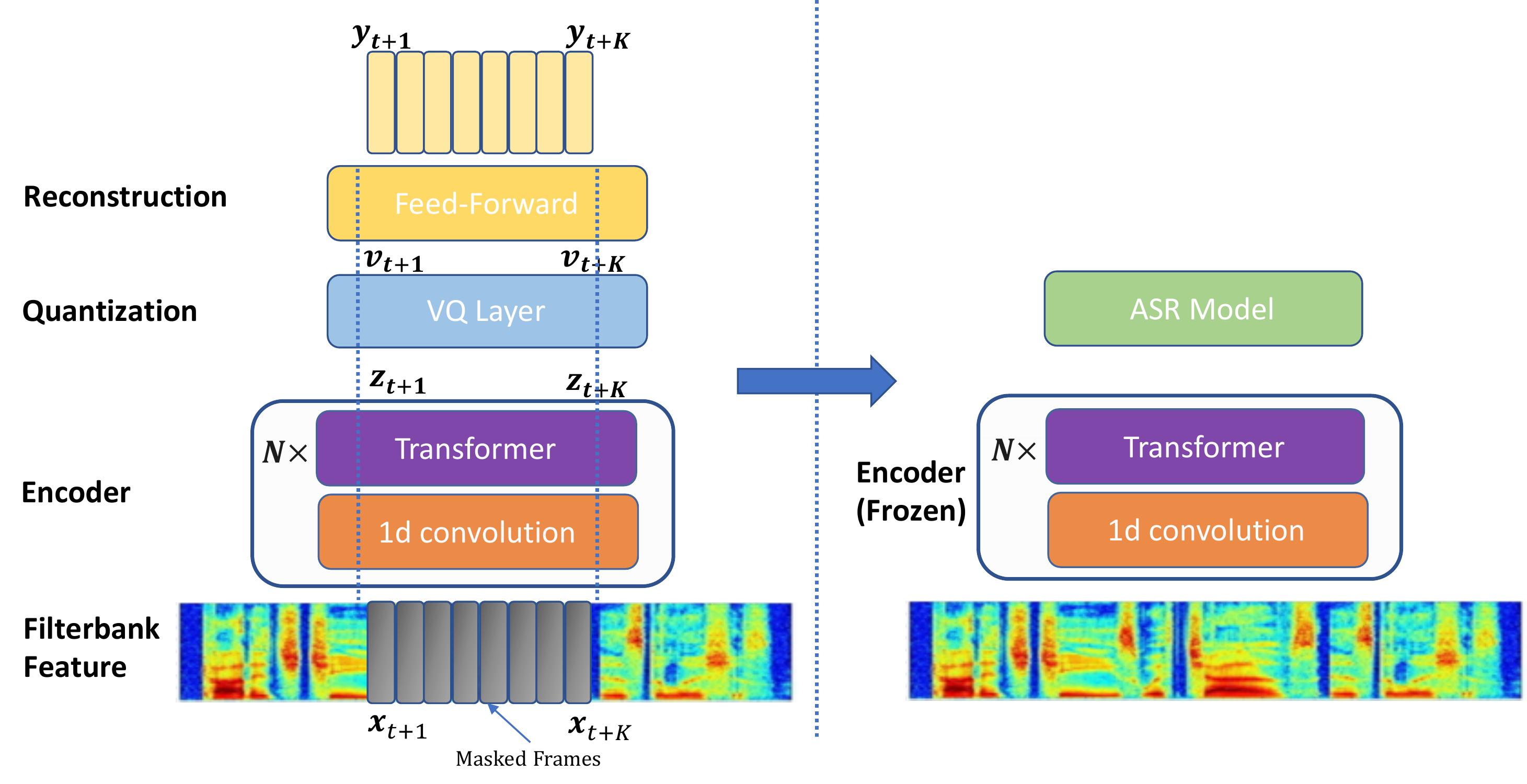}
  \caption{Illustration of DeCoAR 2.0 framework. The left side shows the architecture of speech representation model using unlabeled data. The right side shows an example on using labeled data with the learned speech representation. Note that the quantization and reconstruction modules are removed, and a frozen encoder is attached to a downstream ASR model (such as an acoustic model in hybrid-based system, or an end-to-end ASR system). Only the parameters in ASR model block are trained. }
  \label{fig:model_view}
\end{figure*}
\section{Proposed Framework}
\label{sec:approach}

DeCoAR 2.0 is a follow-up work based on DeCoAR, and we take inspirations of recent advancement in natural language and speech representation learning.  The left figure in Figure~\ref{fig:model_view} illustrates the proposed DeCoAR 2.0 architecture. The model consists of three modules. The first module is the \textbf{encoder network} that maps input masked acoustic frames $\mathbf{x}_{t+1},\cdots,\mathbf{x}_{t+K}$ into a latent representation $\mathbf{z}_{t+1},\cdots,\mathbf{z}_{t+K}$ via multiple Transformer blocks. The second module is the \textbf{vector quantization network} that maps latent representation $\mathbf{z}_{t+1},\cdots,\mathbf{z}_{t+K}$ to a new quantized representation $\mathbf{v}_{t+1},\cdots,\mathbf{v}_{t+K}$. The last module, \textbf{reconstruction network}, takes the quantized representation to a feed-forward network and reconstructs the original input frames as $\mathbf{y}_{t+1},\cdots,\mathbf{y}_{t+K}$. We will describe the design of each module and its training criterion in the following sections.

\subsection{Encoder Module}
We replace forward/backward LSTM with Transformer,  due to its superiority in modeling long context \cite{vaswani2017attention,wang2020transformer}. While RNN/LSTM can model long context in theory, the Transformer achieves better performance thanks to its multi-head attention mechanism that captures the relationship for any arbitrary pair of samples in a long input sequence. In our encoder, we use a 1D convolutional layer with kernel size of 256 and 16 groups. This performs an implicit relative positional encoding as pointed out in \cite{mohamed2019transformers}. The convolution is followed by Gaussian Error Linear Unit (GELU) and layer normalization. The output is then fed into the deep transformer encoder network and produce a sequence of hidden vectors $\mathbf{Z} = (\mathbf{z}_1,\cdots,\mathbf{z}_T)$. 

In our masking strategy, we mask a proportion of the feature and replace them with a trainable feature vector. We randomly mask the subsequent $K$ consecutive time steps from every sampled index; spans are not overlap and we masked around 40\% frames in total.

\subsection{Quantization Module}

We introduce a quantization module in DeCoAR 2.0 framework. Quantization module takes the latent representation $\mathbf{z}_t$ from encoder module, and map it to a new representation $\mathbf{v}_t$. This is done by selecting one entry from a fixed-size codebook $C=\{c_1, \cdots, c_V\}$, where $V$ is the size of the codebook, and apply a linear transformation to obtain $\mathbf{v}_t$. Selecting an entry in a discrete cookbook is not differentiable. To mitigate the problem, we use the Gumbel-Softmax loss with reparameterization trick.  In line with VQ-wav2vec \cite{baevski2019vq}, wav2vec 2.0 \cite{baevski2020wav2vec} and VQ-APC \cite{chung2020vector}, we use the straight-through Gumbel-Softmax estimator~\cite{jang2017categorical}. 

In our quantization module, we use multiple codebooks \cite{baevski2020wav2vec} to obtain quantized representations.  Formally, given the latent representation $\mathbf{z}$ from the encoder module, a set of codebooks $C_1, \cdots, C_G$ where $G$ is the number of codebooks, $V$ entries in each codebook, we select one variable from each codebook and stack the resulting vectors followed by a linear transformation to obtain new representation $\mathbf{v}$. In order to train which entry to select, we map the encoder output $\mathbf{z}$ to logits $\mathbf{l}\in\mathcal{R}^{G\times V}$ via a linear layer, and the probability of selecting the $j$-th code in $g$-th codebook is defined as follows:
\begin{equation}\label{eq:gumble}
  \begin{aligned}
p_{g, j}=\frac{\exp(\mathbf{l}_{g,j}+n_j)/\tau}{\sum_{k=1}^{V}\exp(\mathbf{l}_{g,k}+n_k)/\tau}
\end{aligned}
\end{equation}
where $\tau> 0$ is the softmax temperature, $n=-\ln(-\ln(u))$ and $u$ are uniformly sampled from $\mathcal{U}(0,1)$. In inference, the index with largest value in logits $\mathbf{l}$ is selected from each codebook.

\subsection{Training Objective}

The training objective consists of two parts. The first objective is the \textbf{reconstruction loss}. We use $\ell_1$ loss between an acoustic feature vectors $\mathbf{X}$ at time $t$ and a reconstruction $\mathbf{Y}$  predicted at time $t$ for all masked indices $t$, defined as $\mathcal{L}_{recon}=\sum_t|\mathbf{x}_{t} - \mathbf{y}_t|$. We use $\ell_1$ loss as it is less sensitive to outliers.

Since vector quantization layers are known to significantly disrupt model training, we apply the \textbf{diversity loss} proposed in wav2vec 2.0 \cite{baevski2020wav2vec} to encourage the equal use of all entries in each codebook. Diversity loss maximizes the entropy of the averaged softmax distribution over the entries for each codebook in each mini-batch. Formally, the \textbf{diversity loss} is defined as: 
\begin{equation}\label{eq:div} 
  \begin{aligned}
\mathcal{L}_{div}= \frac{GV-\sum_{g=1}^G \exp(-\sum_{v=1}^V p_{g,j} \log p_{g,j})}{GV} 
\end{aligned}
\end{equation}

Our final training objective is a combination of the two loss functions, weighted by a hyperparameter $\alpha$:
\begin{equation}\label{eq:loss}
  \begin{aligned}
\mathcal{L}=\mathcal{L}_{recon}+\alpha \mathcal{L}_{div}
\end{aligned}
\end{equation}

\subsection{Semi-supervised Speech Recognition with DeCoAR 2.0}

After we have pre-trained the DeCoAR 2.0 model on unlabeled data, we freeze all the parameters in the network. We remove the quantization module and reconstruction module. The representations from the Transformer encoder module are then attached to a downstream ASR system. This ASR system can be either a conventional acoustic model in a hybrid-based ASR system, or an end-to-end speech recognition such as RNN-Transducers \cite{rao2017exploring,li2019improving} or Encoder-Decoder based model \cite{chan2016listen,watanabe2017hybrid}. Note that in our framework, we only train parameters for the downstream ASR model and leave all parameters in the encoder module fixed (i.e. no backpropagation to all layers in encoder module).

\begin{table*}[thb]
    \centering\resizebox{\textwidth}{!}{
   \begin{tabular}{ c | c|cc|cc|cc|cc }
       \toprule
        \multirow{2}{*}{\textbf{Representation}} &
        \multirow{2}{*}{\textbf{Encoder Model}} & \multicolumn{2}{c|}{\textbf{1 hour}} & \multicolumn{2}{c|}{\textbf{10 hours}} & \multicolumn{2}{c}{\textbf{100 hours}} &
        \multicolumn{2}{c}{\textbf{960 hours}}\\
         & &test-clean & test-other & test-clean & test-other & test-clean & test-other &test-clean & test-other\\ 
        \midrule
        filterbank &  - &   50.90 & 78.66 &17.45&  47.18& 9.36 & 30.20  &5.82 &14.50\\
        wav2vec 2.0 \cite{baevski2020wav2vec} & 12 Transformer &  13.63 &29.97 &   5.63& 13.39 & 5.10  & 11.94 &-&- \\
        VQ-APC \cite{chung2020vector}    & 3 uni-GRU                &28.66 & 61.12  & 12.38& 32.28 & 7.42  & 23.38  &-&-  \\
        DeCoAR  \cite{ling2020deep}                    & 4 bi-LSTM &  17.93 &38.38 &  10.40&    27.41&  6.10&17.43  &-&- \\
        DeCoAR 2.0                & 12 Transformer & 13.75 & 29.13 &  5.43& 13.27 & 5.02  & 12.07  &-&-\\
        \bottomrule
    \end{tabular}}
    \caption{Semi-supervised LibriSpeech results. }
    \label{table:LibriSpeech}
\end{table*}

\section{Experimental Setup and Results}
\label{sec:expt}

Our experiments were conducted on the publicly available LibriSpeech dataset. To simulate different SSL scenarios, we varied the labeled data size from 1-hour, 10-hour, up to 100-hour. The 100-hr dataset is based on \textit{train-clean-100} split, and the 1-hr/10-hr subsets are randomly selected from it. 

\subsection{Pretrain DeCoAR 2.0 Model using Unlabeled Data}

To train the DeCoAR 2.0 model, we used the entire 960 hours of LibriSpeech dataset as unlabeled data. We followed the conventional frontend feature extraction, and used a 80-dimensional log-mel filterbank features, which were extracted with a 25ms sliding window at a 10ms frame rate. The features were normalized via mean subtraction and variance normalization on a per-speaker basis. 

For the encoder network in DeCoAR 2.0, we used 12 Transformer blocks, each consists of a multi-head self-attention sublayer followed by a feed forward sublayer. For fair comparison, we set the model dimension to 768, the inner dimension in feed forward sublayer to 3072, with 8 attention heads as used in wav2vec 2.0 base model. The slice size $K$ was set to 20. We optimized the network with Adam and used learning rate warm-up for the first 32000 updates to a peak of 0.0003, and then linearly decayed it. We grouped the input sequences by length with a batch size of 128 (we chopped the maximum length to 15 seconds), and trained the models on 16 GPUs for 150 epochs. The Gumbel softmax temperature $\tau$ is annealed from 2 to a minimum of 0.5 by a factor of $0.999995$ at every update. We use weight $\alpha=0.1$ for the diversity loss and we set $G=2$ and $V=320$ for the quantization module.

\subsection{Semi-supervised Speech Recognition Experimental Results}

We trained acoustic models using CTC loss on labeled data as downstream tasks. Unlike conventional HMM-based hybrid ASR, training acoustic model with CTC loss gets rid of the need to prepare frame-wise alignments and other tedious processes such as preparing state-tying trees.  The total size of CTC labels were 71 phonemes derived from CMU lexicon, plus one blank symbol. For decoding, we used WFST-based decoding using EESEN \cite{miao2015eesen}. CTC labels, lexicons and a 4-gram language model for LibriSpeech were composed into a WFST-based decoding graph. We set the acoustic model scale to $1.0$, and the blank symbol prior scale to $0.3$. We used \textit{dev-clean} for validation and \textit{test-clean}, \textit{test-other} for evaluation.  

We trained different ASR systems for comparison, using different acoustic representations, including wav2vec 2.0 features \cite{baevski2020wav2vec}, VQ-APC features \cite{chung2020vector}, our previously proposed DeCoAR features \cite{ling2020deep}, DeCoAR 2.0 features as proposed in this work. For wav2vec 2.0 features \cite{schneider2019wav2vec}, we obtained 768-dimensional representations from the wav2vec 2.0 base model downloaded from\footnote{https://github.com/pytorch/fairseq/tree/master/examples/wav2vec}, which was pre-trained on 960-hour LibriSpeech data with contrastive loss and had the exactly same encoding network as ours. For VQ-APC features, we trained a VQ-APC model using the official code\footnote{https://github.com/iamyuanchung/VQ-APC} provided by the authors on 960-hour LibriSpeech. We obtained 512-dimensional VP-APC representations as input features. DeCoAR and DeCoAR 2.0 have dimensionality of 2048 and 768, respectively. For all systems trained on learned speech representations, the downstream ASR model are 2 layers of bidirectional LSTMs with CTC loss. In line with our previous work \cite{ling2020deep}, we also train purely supervised systems using conventional filterbank features. These models are trained using 6 layers of bidirectional LSTMs with CTC loss. We also trained a purely supervised system using the entire 960-hour dataset uisng filterbank features as a baseline.

Table \ref{table:LibriSpeech} shows the results on semi-supervised LibriSpeech experiments. We conducted our semi-supervised experiments using 1 hour, 10 hours, and 100 hours of training data. Our proposed approach significantly outperforms the pure supervised filterbank baselines. In particular, under extremely data-sparse conditions, the proposed DeCoAR 2.0 methods achieved highly competitive performance, with a WER of 5.43\%/13.27\% for \textit{test-clean}/\textit{test-other} using 10 hours of labeled data, and a WER of 13.75\%/29.13\% for \textit{test-clean}/\textit{test-other} using only 1 hour of labeled data. One notable observation is that using 10 hours of labeled data can already outperform the system trained on the full 960-hour data with filterbank features by 6.7\%/8.5\% relative WER improvements on \textit{test-clean}/\textit{test-other}.  

Among different speech representations, wav2vec 2.0 and DeCoAR 2.0 performed favorably compared to VQ-APC and DeCoAR.  DeCoAR 2.0 is comparable to wav2vec 2.0 in all different SSL conditions as well. 
It is worth noting that we did not perform fine-tuning for all representation learning layers as these models were trained in different stacks. We are interested in gauging the performance comparison by directly using the resulting speech representations produced from different pre-trained speech representation models. 




We conduct an ablation study to investigate the effect of inserting VQ layer in DeCoAR 2.0 in Table~\ref{table:vq_layer}, and confirm the VQ module is beneficial for ASR tasks. We hypothesize that vector quantization forces the DeCoAR model to reduce the model capacity and focus more on informative factors such as linguistic/phonetic unit discovery and less so on other factors such as speaker traits, acoustic condition. 


\begin{table}[thb]
    \centering
    \ninept
    \begin{tabular}{ c | cc  }
        \toprule
        {\textbf{VQ}} & {\textbf{test clean} } & {\textbf{test other} } \\
        \midrule
        \xmark &  6.29 & 18.54 \\
        \cmark & 5.43  & 13.27\\
        \bottomrule
    \end{tabular}
    \caption{Ablation on the effect of using VQ layer on the Librispeech 10 hours SSL experiment.}
    \label{table:vq_layer}
\end{table}

\section{Conclusion}
\label{sec:conclusion}

In this paper, we present vector quantized \textbf{De}ep \textbf{Co}ntextualized \textbf{A}coustic \textbf{R}epresentation (DeCoAR 2.0), an improved speech representation learning approach based on DeCoAR and vector quantization.  DeCoAR 2.0 has multiple modification over the its predecessor, with a deep Transformer as encoding block, and the addition of a vector quantization module before reconstruction module. In extreme data-limited semi-supervised conditions, we observe that using 10 hours of labeled data with DeCoAR 2.0 achieved performance on par with the system trained on 960 hours of conventional filterbank features. DeCoAR 2.0 also performed comparably to wav2vec 2.0 in all different semi-supervised scenarios. Future work includes exploring the efficacy of representation learning in real world data including noisy and adverse conditions, and extension to neural transducers \cite{rao2017exploring,li2019improving} and other end-to-end ASR systems as downstream tasks. 


\clearpage

\footnotesize
\bibliographystyle{IEEEbib}
\bibliography{refs}

\end{document}